\documentclass[twocolumn,letterpaper,aps,prd,superscriptaddress,showpacs,nofootinbib,floatfix]{revtex4}

\usepackage{graphicx}	

\newcommand{\ua}{\uparrow}
\newcommand{\da}{\downarrow}

\begin{document}


\title{ Inclusive cross section and single transverse spin asymmetry
for very forward neutron production in polarized $p$$+$$p$ collisions at $\sqrt{s}$ = 200 GeV }

\newcommand{\abilene}{Abilene Christian University, Abilene, Texas 79699, USA}
\newcommand{\acadsin}{Institute of Physics, Academia Sinica, Taipei 11529, Taiwan}
\newcommand{\banaras}{Department of Physics, Banaras Hindu University, Varanasi 221005, India}
\newcommand{\barc}{Bhabha Atomic Research Centre, Bombay 400 085, India}
\newcommand{\bnlcoll}{Collider-Accelerator Department, Brookhaven National Laboratory, Upton, New York 11973-5000, USA}
\newcommand{\bnlphys}{Physics Department, Brookhaven National Laboratory, Upton, New York 11973-5000, USA}
\newcommand{\caucr}{University of California - Riverside, Riverside, California 92521, USA}
\newcommand{\charlesczech}{Charles University, Ovocn\'{y} trh 5, Praha 1, 116 36, Prague, Czech Republic}
\newcommand{\ciae}{Science and Technology on Nuclear Data Laboratory, China Institute of Atomic Energy, Beijing 102413, P.~R.~China}
\newcommand{\cns}{Center for Nuclear Study, Graduate School of Science, University of Tokyo, 7-3-1 Hongo, Bunkyo, Tokyo 113-0033, Japan}
\newcommand{\colorado}{University of Colorado, Boulder, Colorado 80309, USA}
\newcommand{\columbia}{Columbia University, New York, New York 10027 and Nevis Laboratories, Irvington, New York 10533, USA}
\newcommand{\czechtech}{Czech Technical University, Zikova 4, 166 36 Prague 6, Czech Republic}
\newcommand{\dapnia}{Dapnia, CEA Saclay, F-91191, Gif-sur-Yvette, France}
\newcommand{\debrecen}{Debrecen University, H-4010 Debrecen, Egyetem t{\'e}r 1, Hungary}
\newcommand{\elte}{ELTE, E{\"o}tv{\"o}s Lor{\'a}nd University, H - 1117 Budapest, P{\'a}zm{\'a}ny P. s. 1/A, Hungary}
\newcommand{\fit}{Florida Institute of Technology, Melbourne, Florida 32901, USA}
\newcommand{\fsu}{Florida State University, Tallahassee, Florida 32306, USA}
\newcommand{\gsu}{Georgia State University, Atlanta, Georgia 30303, USA}
\newcommand{\hiroshima}{Hiroshima University, Kagamiyama, Higashi-Hiroshima 739-8526, Japan}
\newcommand{\ihepprot}{IHEP Protvino, State Research Center of Russian Federation, Institute for High Energy Physics, Protvino, 142281, Russia}
\newcommand{\illuiuc}{University of Illinois at Urbana-Champaign, Urbana, Illinois 61801, USA}
\newcommand{\inrras}{Institute for Nuclear Research of the Russian Academy of Sciences, prospekt 60-letiya Oktyabrya 7a, Moscow 117312, Russia}
\newcommand{\instpasczech}{Institute of Physics, Academy of Sciences of the Czech Republic, Na Slovance 2, 182 21 Prague 8, Czech Republic}
\newcommand{\isu}{Iowa State University, Ames, Iowa 50011, USA}
\newcommand{\jinrdubna}{Joint Institute for Nuclear Research, 141980 Dubna, Moscow Region, Russia}
\newcommand{\kek}{KEK, High Energy Accelerator Research Organization, Tsukuba, Ibaraki 305-0801, Japan}
\newcommand{\korea}{Korea University, Seoul, 136-701, Korea}
\newcommand{\kurchatov}{Russian Research Center ``Kurchatov Institute", Moscow, 123098 Russia}
\newcommand{\kyoto}{Kyoto University, Kyoto 606-8502, Japan}
\newcommand{\labllr}{Laboratoire Leprince-Ringuet, Ecole Polytechnique, CNRS-IN2P3, Route de Saclay, F-91128, Palaiseau, France}
\newcommand{\lawllnl}{Lawrence Livermore National Laboratory, Livermore, California 94550, USA}
\newcommand{\losalamos}{Los Alamos National Laboratory, Los Alamos, New Mexico 87545, USA}
\newcommand{\lpc}{LPC, Universit{\'e} Blaise Pascal, CNRS-IN2P3, Clermont-Fd, 63177 Aubiere Cedex, France}
\newcommand{\lund}{Department of Physics, Lund University, Box 118, SE-221 00 Lund, Sweden}
\newcommand{\mass}{Department of Physics, University of Massachusetts, Amherst, Massachusetts 01003-9337, USA }
\newcommand{\muenster}{Institut f\"ur Kernphysik, University of Muenster, D-48149 Muenster, Germany}
\newcommand{\muhlenberg}{Muhlenberg College, Allentown, Pennsylvania 18104-5586, USA}
\newcommand{\myongji}{Myongji University, Yongin, Kyonggido 449-728, Korea}
\newcommand{\nagasaki}{Nagasaki Institute of Applied Science, Nagasaki-shi, Nagasaki 851-0193, Japan}
\newcommand{\newmex}{University of New Mexico, Albuquerque, New Mexico 87131, USA }
\newcommand{\nmsu}{New Mexico State University, Las Cruces, New Mexico 88003, USA}
\newcommand{\ornl}{Oak Ridge National Laboratory, Oak Ridge, Tennessee 37831, USA}
\newcommand{\orsay}{IPN-Orsay, Universite Paris Sud, CNRS-IN2P3, BP1, F-91406, Orsay, France}
\newcommand{\peking}{Peking University, Beijing 100871, P.~R.~China}
\newcommand{\pnpi}{PNPI, Petersburg Nuclear Physics Institute, Gatchina, Leningrad region, 188300, Russia}
\newcommand{\riken}{RIKEN Nishina Center for Accelerator-Based Science, Wako, Saitama 351-0198, Japan}
\newcommand{\rikjrbrc}{RIKEN BNL Research Center, Brookhaven National Laboratory, Upton, New York 11973-5000, USA}
\newcommand{\rikkyo}{Physics Department, Rikkyo University, 3-34-1 Nishi-Ikebukuro, Toshima, Tokyo 171-8501, Japan}
\newcommand{\saispbstu}{Saint Petersburg State Polytechnic University, St. Petersburg, 195251 Russia}
\newcommand{\saopaulo}{Universidade de S{\~a}o Paulo, Instituto de F\'{\i}sica, Caixa Postal 66318, S{\~a}o Paulo CEP05315-970, Brazil}
\newcommand{\seoulnat}{Seoul National University, Seoul, Korea}
\newcommand{\stonybrkc}{Chemistry Department, Stony Brook University, SUNY, Stony Brook, New York 11794-3400, USA}
\newcommand{\stonycrkp}{Department of Physics and Astronomy, Stony Brook University, SUNY, Stony Brook, New York 11794-3400, USA}
\newcommand{\subatech}{SUBATECH (Ecole des Mines de Nantes, CNRS-IN2P3, Universit{\'e} de Nantes) BP 20722 - 44307, Nantes, France}
\newcommand{\tenn}{University of Tennessee, Knoxville, Tennessee 37996, USA}
\newcommand{\titech}{Department of Physics, Tokyo Institute of Technology, Oh-okayama, Meguro, Tokyo 152-8551, Japan}
\newcommand{\tsukuba}{Institute of Physics, University of Tsukuba, Tsukuba, Ibaraki 305, Japan}
\newcommand{\vandy}{Vanderbilt University, Nashville, Tennessee 37235, USA}
\newcommand{\waseda}{Waseda University, Advanced Research Institute for Science and Engineering, 17 Kikui-cho, Shinjuku-ku, Tokyo 162-0044, Japan}
\newcommand{\weizmann}{Weizmann Institute, Rehovot 76100, Israel}
\newcommand{\wigner}{Institute for Particle and Nuclear Physics, Wigner Research Centre for Physics, Hungarian Academy of Sciences (Wigner RCP, RMKI) H-1525 Budapest 114, POBox 49, Budapest, Hungary}
\newcommand{\yonsei}{Yonsei University, IPAP, Seoul 120-749, Korea}
\affiliation{\abilene}
\affiliation{\acadsin}
\affiliation{\banaras}
\affiliation{\barc}
\affiliation{\bnlcoll}
\affiliation{\bnlphys}
\affiliation{\caucr}
\affiliation{\charlesczech}
\affiliation{\ciae}
\affiliation{\cns}
\affiliation{\colorado}
\affiliation{\columbia}
\affiliation{\czechtech}
\affiliation{\dapnia}
\affiliation{\debrecen}
\affiliation{\elte}
\affiliation{\fit}
\affiliation{\fsu}
\affiliation{\gsu}
\affiliation{\hiroshima}
\affiliation{\ihepprot}
\affiliation{\illuiuc}
\affiliation{\inrras}
\affiliation{\instpasczech}
\affiliation{\isu}
\affiliation{\jinrdubna}
\affiliation{\kek}
\affiliation{\korea}
\affiliation{\kurchatov}
\affiliation{\kyoto}
\affiliation{\labllr}
\affiliation{\lawllnl}
\affiliation{\losalamos}
\affiliation{\lpc}
\affiliation{\lund}
\affiliation{\mass}
\affiliation{\muenster}
\affiliation{\muhlenberg}
\affiliation{\myongji}
\affiliation{\nagasaki}
\affiliation{\newmex}
\affiliation{\nmsu}
\affiliation{\ornl}
\affiliation{\orsay}
\affiliation{\peking}
\affiliation{\pnpi}
\affiliation{\riken}
\affiliation{\rikjrbrc}
\affiliation{\rikkyo}
\affiliation{\saispbstu}
\affiliation{\saopaulo}
\affiliation{\seoulnat}
\affiliation{\stonybrkc}
\affiliation{\stonycrkp}
\affiliation{\subatech}
\affiliation{\tenn}
\affiliation{\titech}
\affiliation{\tsukuba}
\affiliation{\vandy}
\affiliation{\waseda}
\affiliation{\weizmann}
\affiliation{\wigner}
\affiliation{\yonsei}
\author{A.~Adare} \affiliation{\colorado}
\author{S.~Afanasiev} \affiliation{\jinrdubna}
\author{C.~Aidala} \affiliation{\mass}
\author{N.N.~Ajitanand} \affiliation{\stonybrkc}
\author{Y.~Akiba} \affiliation{\riken} \affiliation{\rikjrbrc}
\author{H.~Al-Bataineh} \affiliation{\nmsu}
\author{J.~Alexander} \affiliation{\stonybrkc}
\author{K.~Aoki} \affiliation{\kyoto} \affiliation{\riken}
\author{L.~Aphecetche} \affiliation{\subatech}
\author{J.~Asai} \affiliation{\riken}
% ...
\collaboration{PHENIX Collaboration} \noaffiliation
   
\date{\today}
   
\begin{abstract}

The energy dependence of the single-transverse-spin asymmetry, 
$A_N$, and the cross section for neutron production at very forward 
angles were measured in the PHENIX experiment at RHIC for polarized 
$p$$+$$p$ collisions at $\sqrt{s}$=200 GeV. The neutrons were 
observed in forward detectors covering an angular range of up to 
2.2 mrad. We report results for neutrons with momentum fraction of 
$x_F$=0.45 to 1.0. The energy dependence of the measured cross 
sections were consistent with $x_F$ scaling, compared to 
measurements by an ISR experiment which measured neutron production 
in unpolarized $p$$+$$p$ collisions at $\sqrt{s}$=30.6--62.7 GeV. 
The cross sections for large $x_F$ neutron production for $p$$+$$p$ 
collisions, as well as those in $e+p$ collisions measured at HERA, 
are described by a pion exchange mechanism. The observed forward 
neutron asymmetries were large, reaching $A_N=-0.08\pm0.02$ for 
$x_F$=0.8; the measured backward asymmetries, for negative $x_F$, 
were consistent with zero. The observed asymmetry for forward 
neutron production is discussed within the pion exchange framework, 
with interference between the spin-flip amplitude due to the pion 
exchange and nonflip amplitudes from all Reggeon exchanges.  
Within the pion exchange description, the measured neutron 
asymmetry is sensitive to the contribution of other Reggeon 
exchanges even for small amplitudes.
   
\end{abstract}
   
\pacs{13.85.Ni,13.88.+e,14.20.Dh,25.75.Dw}
   
\maketitle
   
 
 
\section{Introduction}

With the first polarized $p$+$p$ collisions at $\sqrt{s}$ = 200 GeV 
at the Relativistic Heavy Ion Collider (RHIC), a large single 
transverse spin asymmetry ($A_N$) for neutron production in very 
forward kinematics was discovered by a polarimeter development 
experiment \cite{Fukao:2006vd}. That experiment was designed to 
measure the asymmetry for very forward photons and used an 
electromagnetic calorimeter. The calorimeter was used to identify 
neutrons, originally to remove them from the photon data, when a 
large asymmetry was observed in forward neutrons. The neutron 
energy resolution was coarse, so no cross section measurement was 
reported. The discovery inspired the PHENIX experiment to use 
existing very forward hadronic calorimeters, with additional shower 
maximum detectors, to measure the neutron transverse asymmetry at 
the PHENIX interaction point at RHIC with a significantly better 
neutron energy resolution. Here we report the first measurements of 
very forward inclusive and semi-inclusive neutron production cross 
sections at $\sqrt{s}$ = 200 GeV and measurements of $A_N$ for 
forward and backward production with improved neutron energy 
resolution. The $A_N$ is a left--right asymmetry written as:

\begin{equation}
  A_N = \frac{d\sigma^\ua -\ d\sigma^\da}{d\sigma^\ua +\ 
d\sigma^\da}
\end{equation}

\noindent for yields observed to the left when facing along the 
polarized proton's momentum vector, where $d\sigma^\ua$ 
($d\sigma^\da$) is the production cross section when the protons 
are polarized up (down). The $A_N$ with cross section measurements for 
higher energy $p$$+$$p$ collisions provide qualitatively new 
information toward an understanding of the production mechanism.

\begin{figure}[htb]
\includegraphics[width=0.8\linewidth]{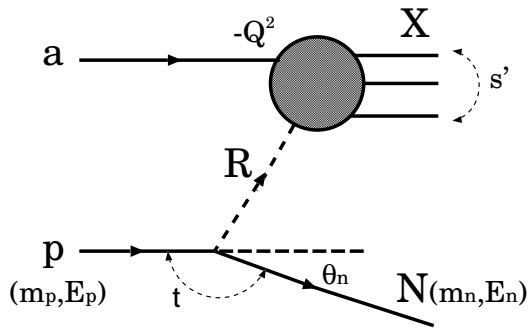}
\caption{
A schematic diagram of neutron production, $pa \rightarrow nX$, for 
the Reggeon exchange model shown with Lorentz invariant variables 
$s', Q^2$ and $t$. ``$a$'' is a proton or positron for $p$$+$$p$ or 
$e^{+}p$ reactions. $R$ indicates a Regge trajectory with isospin 
odd such as $\pi$, $\rho$, $a_{2}$ and Pomeron-$\pi$ in the Regge 
theory. For pion exchange, $R$ = $\pi$.
}
\label{fig_i_1}
\end{figure}

Cross sections of inclusive neutron production in unpolarized 
$p$$+$$p$ collisions were measured at the ISR from $\sqrt{s}$ = 
30.6 to 62.7 GeV \cite{Engler:1974nz,Flauger:1976ju}.  These cross 
sections have been described using One Pion Exchange (OPE) models 
\cite{Capella:1975qk,Kopeliovich:1996iw, 
Nikolaev:1997cn,Nikolaev:1998se,D'Alesio:1998bf,Kaidalov:2006cw, 
Bunyatyan:2006vy}.  In OPE, the incoming proton emits a pion which 
scatters on the other proton as shown in Figure \ref{fig_i_1}. 
Kinematics of the neutron are characterized by two variables, $x_F$ 
and $p_T$ defined by,

\begin{eqnarray}
  x_{F} &=& p_{L}/p_{L(max)}
          = E_{n}\cos{\theta_{n}}/E_{p} \sim E_{n}/E_{p}, \label{eq_i_2} \\
  p_{T} &=& E_{n}\sin{\theta_{n}} \sim x_{F}E_{p}\theta_{n}. \label{eq_i_3}
\end{eqnarray} 

\noindent where $p_L$ is the momentum component of the neutron in 
the proton-beam direction, $E_n$ and $E_p$ are energies of the 
neutron and the proton beam, and $\theta_n$ is the polar angle of 
the neutron from the beam direction which is very small 
($\sim$mrad) for forward neutron production. The measured cross 
section showed a peak around $x_F \sim 0.8$ and was found to have 
almost no $\sqrt{s}$ dependence. OPE models gave a reasonable 
description of the data.

OPE models were also used to describe proton and photon induced 
production of neutrons measured at the HERA $e+p$ collider 
\cite{Chekanov:2007tv,Aaron:2010ze}.  These measurements probe the 
pion structure function at small $x$. The NA49 collaboration also 
published the cross section for forward neutron production for 
$p$$+$$p$ collisions at $\sqrt{s}$ = 17.3 GeV 
\cite{Anticic:2009wd}. They compared the result with those from ISR 
and HERA and found they did not agree.

The neutron asymmetry provides a new tool to probe the production 
mechanism. For the OPE model, $A_N$ arises from an interference 
between spin-flip and spin-nonflip amplitudes. Since the 
pion-exchange amplitude is fully spin-flip, the asymmetry is 
sensitive to other Reggeon exchange amplitudes which are 
spin-nonflip, even for small amplitudes.

This paper presents the $x_{F}$ dependence of cross sections, 
inclusive and semi-inclusive (with a beam-beam interaction 
requirement), and $A_N$ for very forward and very backward neutron 
production in polarized $p$$+$$p$ collisions at $\sqrt{s}$ = 200 
GeV.

\section{Experimental setup}

\subsection{Detector apparatus}\label{sec_e}

\begin{figure}[htb]
\includegraphics[width=0.32\linewidth,angle=270]{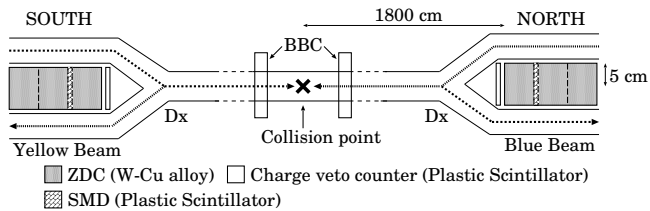}
\caption{
A plan view of the experimental setup  at PHENIX (not to scale).
Shown are the principal components for the neutron physics.
Charged veto counters are in front of ZDCs, and the SMDs are between 
the first and second ZDC modules. 
}
\label{fig_e_1}
\end{figure}

A plan view of the experimental setup for very forward neutron 
measurement at PHENIX \cite{Adcox:2003zm} is shown in 
Fig.~\ref{fig_e_1}. The RHIC polarized proton beams were vertically 
polarized. Each collider ring of RHIC was filled with up to 111 
bunches in a 120 bunch pattern, spaced 106 ns apart, with 
predetermined patterns of polarization signs for the bunches. The 
colliding beam rotating clockwise when viewed from above is 
referred to as the ``Blue beam'' and the beam rotating 
counterclockwise, the ``Yellow beam''.

Neutrons were measured by a Zero-Degree Calorimeter (ZDC) \cite{Adler:2000bd} 
with a position-sensitive Shower-Maximum Detector (SMD). 
One ZDC module is composed of Cu-W alloy absorbers with PMMA-based
communication grade optical fibers, and corresponds to 1.7 nuclear 
interaction lengths.
A single photomultiplier collects \v{C}erenkov light via optical fibers.
Three ZDC modules are located in series (5.1 nuclear interaction lengths) 
at $\pm$1800 cm away from the collision point, covering
10 cm in the transverse plane.

The SMD comprises $x$-$y$ scintillator strip hodoscopes and is 
inserted between the first and second ZDC modules (see Fig.~5 of 
\cite{Adler:2000bd}) at approximately the depth of the maximum 
of the hadronic shower.
The $x$-coordinate (horizontal) is given by 7 scintillator strips of 15 mm 
width, while the $y$-coordinate (vertical) is given by 8 strips of 20 mm 
width, tilted by 45 degrees.

The neutron position can be reconstructed from the energy deposited in 
scintillators with the centroid method. 
We calculated the centroid: 
 \begin{eqnarray}
   x = \frac{ \sum^{N^{SMD}_{multi.}}_{i} E(i) \cdot x(i) }
{ \sum^{N^{SMD}_{multi.}}_{i} E(i) }, \label{eq_e_1}
 \end{eqnarray}
where $E(i)$ and $x(i)$ are the energy deposit and the position of the 
$i$-th scintillator, respectively. 
The number of scintillators with pulse height above the Minimum Ionization 
Particle (MIP) peak is shown as $N^{SMD}_{multi.}$ which is defined as 
the SMD multiplicity. 

Detectors are located downstream of the RHIC dipole (DX) magnet, so 
that collision-related charged particles are swept out. A forward 
scintillation counter, with dimensions 10$\times$12 cm, was 
installed in front of the ZDC to remove charged particle 
backgrounds from other sources. In this analysis, we used only the 
south ZDC detector, which is facing the Yellow beam.

As a beam luminosity monitor, Beam Beam Counters (BBCs) are used.
The BBC comprises 64 photomultiplier tubes 
and 3 cm thick quartz \v{C}erenkov radiators.
The two BBCs are mounted around the beam pipe $\pm$144 cm away from the 
collision point
which cover $\pm$(3.0--3.9) in pseudorapidity and $2\pi$ in azimuth. 
 
The neutron data were collected in 2006 with two triggers.
One is the ZDC trigger for neutron inclusive measurements, requiring an energy 
deposit in the south ZDC greater than 5 GeV.
The other trigger was a ZDC$\otimes$BBC trigger, a coincidence trigger of the 
ZDC trigger with BBC hits which are defined as one or more charged particles 
in both of the BBC detectors.
We note that the ZDC trigger was prescaled due to data acquisition limitations.
Therefore, the ZDC trigger samples are significantly smaller than the 
ZDC$\otimes$BBC trigger samples. 
 
\subsection{Detector performance}\label{sec_e2}

In order to evaluate the detector performance, simulation studies were 
performed with {\sc geant}3 with GHEISHA \cite{Brun:1994aa} 
which simulated the response of the prototype ZDC to hadrons well.
A single neutron event generator and 
{\sc pythia}  (version 6.220) \cite{Sjostrand:2000wi} were used to generate
events.
The single neutron event generator simulated neutrons as a function of $x_{F}$ 
and $p_{T}$. 
The $x_F$ distribution which was used for the simulation input was determined 
as a differential cross section, $d\sigma/dx_F$, in the cross section analysis 
(section \ref{sec_ax_1}).
The $p_T$ distribution is difficult to determine by the PHENIX data alone since
the position and energy resolutions are insufficient to adequately determine it, so
the $p_T$ distribution from the ISR result, exp($-$4.8 $p_T$ (GeV/$c$)), was used as
simulation input, assuming $p_T$ scaling from the ISR to the PHENIX energies.
To check the reliability of this assumption, distributions of radial distance from the 
detector center, $r$, for the data and simulation were compared based on the 
relation of $p_T \propto r$ as,
\begin{eqnarray}
  p_T = E_n {\rm sin}\theta_n = E_n \frac{r}{\sqrt{r^2+d^2}} \sim E_n\frac{r}{d},
\end{eqnarray}
where $d$ is the distance from the collision point to the detector, corresponding to 1800 cm, and $r$ is determined for the shower centroid with 
Eq.~(\ref{eq_e_1}).

\begin{figure}[htb]
\includegraphics[width=1.0\linewidth]{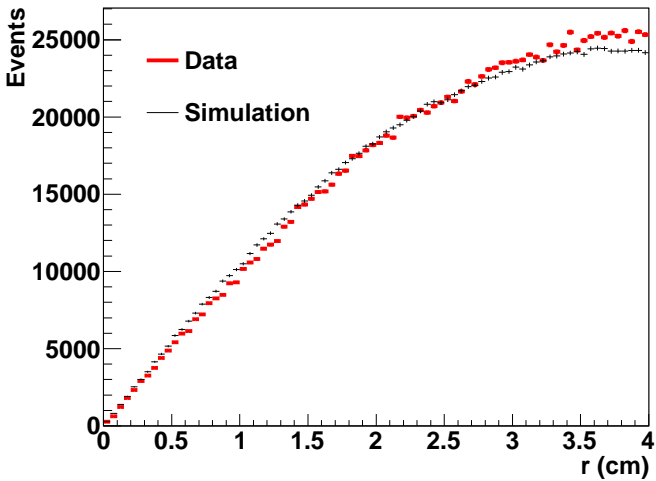}
\caption{(color online)
$r$ distributions for the data and simulation with the exponential $p_T$ 
shape. Distributions agreed within $r$$<$4 cm. 
}
\label{fig_m_2:2}
\end{figure}

The comparison of $r$ distributions with the integration of measured ZDC 
energies 20--120 GeV agreed well as shown in Fig.~\ref{fig_m_2:2}. 

\subsubsection{Performance of the energy measurement}\label{sec_m_3}

The neutron energy measurement with the ZDC was degraded by
a nonlinearity of the photoelectron yield and 
shower leakage out the back and sides of the detector (edge effect).
The ZDC response was studied by simulation with the single neutron event
generator.

\begin{figure}[htb]
\includegraphics[width=1.0\linewidth]{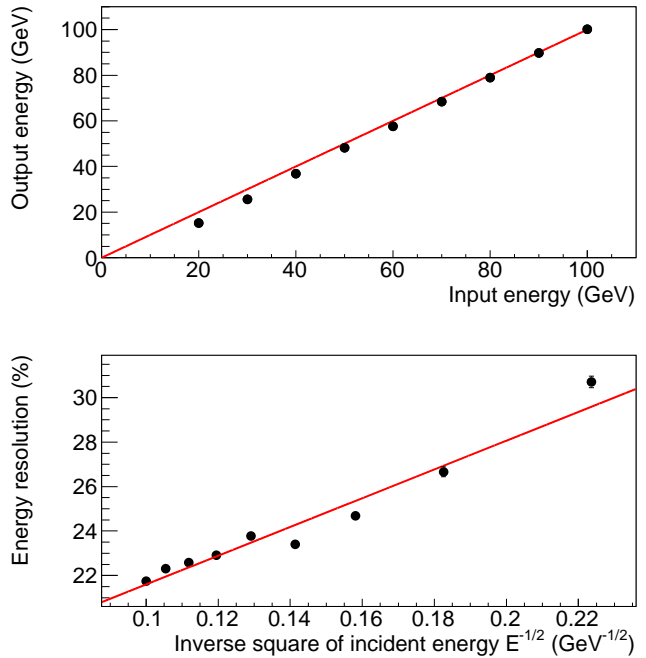}
\caption{(color online)
Top) The mean of output energy as a function of the incident neutron energy
evaluated by the simulation. Solid line indicates a linear response.
Bottom) The energy resolution as a function of 1/$\sqrt{E}$ (GeV$^{-1/2}$).
Solid line shows the fit result; $\Delta E/E$ = 65\%/$\sqrt{E}$ + 15\%.
}
\label{fig_e_1_5}
\end{figure}

The energy linearity and resolution were evaluated from the response
to incident neutrons with energies from 20 to 100 GeV in the simulation. 
The absolute scale was normalized at 100 GeV with the experimental data. 
The ZDC response below 100 GeV exhibits nonlinear
behavior as shown in Fig.~\ref{fig_e_1_5}.
We applied a correction of the nonlinearity to the experimental data 
based on this result. 
We used the difference between the linear and nonlinear response
as a component of the systematic uncertainty in
the determination of the cross section (section \ref{sec_ax_1}).

 As shown in Fig.~\ref{fig_e_1_5}, the energy resolution for 20--100 GeV neutrons 
was described by
\begin{eqnarray}
  \frac{\Delta E}{E} = \frac{65\%}{\sqrt{E\ {\rm (GeV)}}} + 15\%. \label{eq_s_1}
\end{eqnarray}

The absolute scale of the energy measurement was normalized with the 100 GeV single 
neutron peak in heavy ion collisions. 
However, the energy of neutrons from $p$$+$$p$ collisions was below 100 GeV,
so simulation was used to estimate
the detector response for neutron energies in this region.

\begin{figure}[htb]
\includegraphics[width=1.0\linewidth]{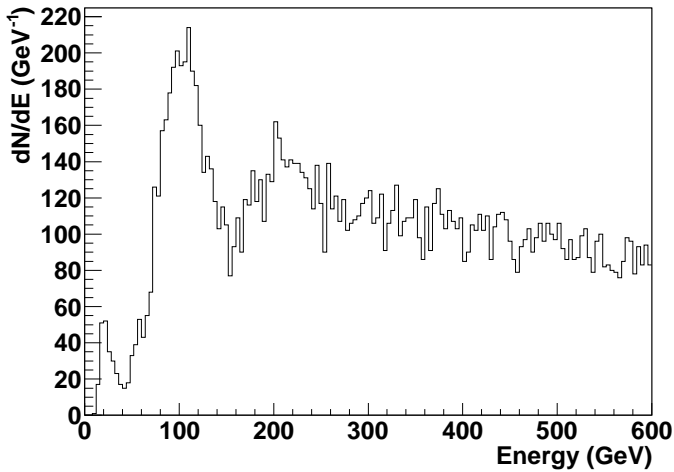}
\caption{
The energy distribution in the ZDC for Cu+Cu collisions at $\sqrt{s_{NN}}$ = 200 GeV.
Peripheral events were selected by requiring BBC inactivity.
}
\label{fig_e_2}
\end{figure}

\begin{figure}[htb]
\includegraphics[width=1.0\linewidth]{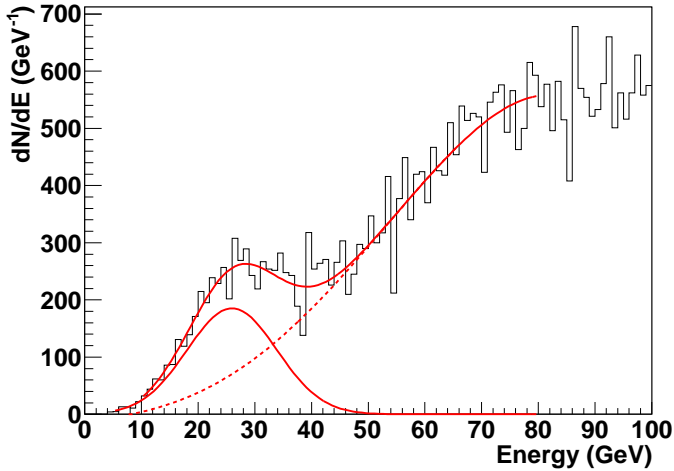}
\caption{(color online)
The energy distribution in the ZDC for Cu+Cu collision at $\sqrt{s_{NN}}$ = 
62.4 GeV. 
The neutron peak position was determined with a Gaussian + polynomial fit.
}
\label{fig_e_3}
\end{figure}

Figure \ref{fig_e_2} shows the absolute energy scale calibrated by observing 
one neutron from peripheral Cu+Cu collisions at $\sqrt{s_{NN}}$ = 200 GeV; 
100 GeV neutrons less than 2 mrad from the beam axis produced
the single neutron peak. 
The energy resolution expected from simulation was about 22\% for the 100 GeV 
neutron and was consistent with the observed width of the single neutron peak as shown 
in Fig.~\ref{fig_e_2}. 
The energy nonlinearity was confirmed by the single neutron peak from Cu+Cu 
collisions at $\sqrt{s_{NN}}$ = 62.4 GeV shown in Fig.~\ref{fig_e_3} which
peaked at 26$\pm$0.3 GeV, consistent 
with nonlinearity indicated by the simulation.

 The edge effect was studied by a prototype ZDC with a 100 GeV proton beam at CERN.
Generally, the measured energy decreased near the edge,
however, nearest the PMT, the measured energy increased.
This was found to be caused by the fibers in the top region which connected to the PMT (see Fig.~5 of \cite{Adler:2000bd}); 
where the shower hit the fibers directly.
The simulation used to study the prototype reproduced this effect.

A residual edge effect was seen in the data at the top and bottom of the detector, so
we chose to apply a fiducial cut to minimize the effect.
According to the simulation, 95--100\% of the incident energy was contained
within $r<$ 3 cm.

\subsubsection{Performance of the position measurement}\label{sec_m_3p}

\begin{figure}[htb]
\includegraphics[width=1.0\linewidth]{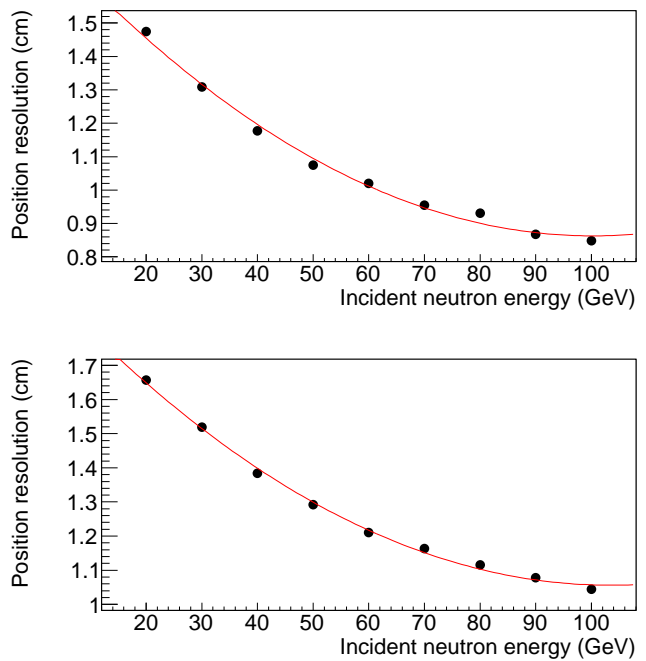}
\caption{(color online)
The position resolution (RMS) as a function of the incident neutron energy for 
$x$ (top) and $y$ (bottom). 
Circles show measured values. 
They were well reproduced by a second order polynomial fit. 
Red lines show the fit results. 
}
\label{fig_m_7}
\end{figure}

 The position resolutions were evaluated by the simulation. 
 Figure \ref{fig_m_7} shows the position resolution (RMS) as a function of the neutron 
incident energy for $x$ (horizontal) and $y$ (vertical) positions. 
 The position resolution was approximately 1 cm for the neutron energy at 100 GeV. 

 Near the edge of the detector, the position measurement is also affected by 
shower leakage. 
 If the incident position was in the edge area, the output position was shifted to the 
detector center due to shower leakage, independent of neutron energy.
 This position shift caused by the edge effect is corrected based on the simulation.

 The reliability of the position measurement was studied by comparing hadron shower 
shapes of the data and  simulation.
 The shower width and highest shower fraction among all scintillators were calculated for 
$x$ and $y$ independently.
 We compared the measured distribution with simulation for each 
SMD multiplicity since the hadron shower shape sensitively depends on the SMD multiplicity.
 The distribution of $y$ was well reproduced by the simulation,
however the distribution of $x$ was not well reproduced, especially for the highest 
shower fraction in high SMD multiplicity events.
 The  systematic uncertainties for the position measurement were estimated by matching 
the highest shower fraction of $x$ by smearing the simulated shower shapes in case 
of the SMD multiplicity = 7, which shows the worst agreement between the data and 
simulation.
After the smearing to match the highest shower fraction, the shower width of 
the simulation also reproduced that of the experimental data. 
The position resolution increased 14\% after the smearing.

\subsubsection{Performance of the neutron identification}\label{sec_m_9}

Events within the detector acceptance in $p$$+$$p$ collisions were
studied with {\sc geant}3 with {\sc pythia}  event generators, and the 
performance of neutron identification and its reliability were evaluated. 

\begin{figure}[htb]
\includegraphics[width=1.0\linewidth]{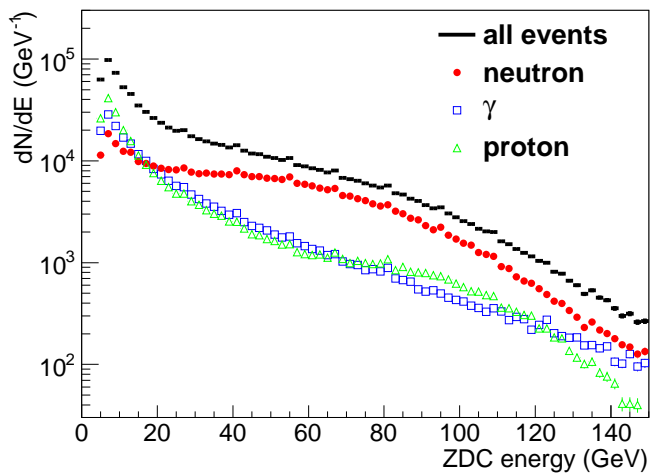}
\caption{(color online)
 Energy distributions in ZDC for neutron, photon and proton.
 The ZDC threshold was set at 5 GeV in the ZDC trigger.
 Events with one particle detected in the ZDC in each $p$$+$$p$ collision are shown.
}
\label{fig_m_19}
\end{figure}

We studied particle species detected in the ZDC with the 5 GeV energy 
threshold which was required for the ZDC trigger (without the BBC 
coincidence requirement). 
In about 92\% of events, only a single particle was detected by the ZDC in 
each $p$$+$$p$ collision, mainly photons, neutrons and protons. 
Energy distributions for these three particles are plotted in 
Fig.~\ref{fig_m_19}. 

 Only neutral particles, photons and neutrons, were expected to be detected with 
the ZDC due to sweeping of charged particles by the DX magnet.
 However scattered protons could hit the DX magnet or beam pipe 
and create a hadronic shower and 
particles from the shower could hit the ZDC.

Most of the photons and neutrons were generated by diffractive and gluon 
scattering processes.
In {\sc pythia}  hard processes, neutrons are generated mainly 
from string fragmentation ($\sim$65\%) and then decay from 
$\Delta^0, \Delta^+, \Delta^-, \Lambda^0$.
The forward photons were generated by decays of $\pi^0$s 
($\sim$91\%) and $\eta$s ($\sim$7\%).
Protons were generated by elastic and diffractive processes. 
Particles depositing less than 20 GeV of energy in the ZDC were predominantly  
photons and protons as shown in Fig.~\ref{fig_m_19}.

Photons are mostly absorbed in the first ZDC module, which is 51 radiation lengths long. 
Thus, photons were removed by requiring energy deposited in the SMD or 
in the second ZDC module. 
In photon rejection with the SMD, more than one scintillator above threshold 
(the SMD multiplicity $\geq$ 2) were required for both $x$ and $y$. 
After applying this cut, the neutron purity was estimated to be 
93.6$\pm$0.3\%. 
In photon rejection with the second ZDC module, energy deposited in 
the second ZDC module above 20 GeV was required. 
After applying this cut, the neutron purity was estimated to be 
93.6$\pm$0.5\%. 
In the analyses of the cross section and the asymmetry, photon rejection 
with the SMD was applied since the position information calculated by 
the SMD was required. 
Rejection with the second ZDC was used for the estimation of 
the rejection efficiency with the SMD which is discussed in section 
\ref{sec_ax_1}.

The charge veto counter was used to reject protons.
A neutron energy above 20 GeV and the charge veto cut 
removed most proton events, as discussed later in this section. 

The main backgrounds after neutron identification are $K^0$s and protons. 
The purities were estimated for neutron energies above 20 GeV. 
In the cross section and the asymmetry analyses, we also required
the acceptance cut and/or a higher energy cut. 
In these cases, the purities improve and are estimated in each analysis 
section. 

In the ISR experiment, the $K^0$ contamination to the neutron measurement 
was estimated from the $K^{\pm}$ measurements \cite{Flauger:1976ju}. 
They obtained 10 \% contamination at $x_F$=0.2 and less than 4 \% at 
$x_F>$0.4. 
The fraction of $K^0$ to neutron in {\sc pythia}  is consistent with the ISR 
result.
We have included no correction for the $K^{0}$ contamination in this 
analysis. 

The proton background is very sensitive to the materials around the ZDC and 
the magnet tuning in the accelerator. 
The systematic uncertainty of proton contamination was estimated by 
the simulation using the measured fraction of charged events 
in the charge veto counter. 
Noise was estimated by the pedestal width of the data and was incorporated 
into the simulation.
For the proton contamination analysis, photon events were removed by 
requiring the second ZDC module cut. 
The fraction of proton events can be estimated as a fraction of charged 
candidates, which are the events with one more MIPs in the charge veto 
counter. 
These fractions were 0.42 and 0.28 for the data and simulation, 
respectively. 
Proton events in the experimental data were about 1.5 times more frequent 
than that of the simulation. 
We ascribe the difference to beam conditions that cause
interactions with materials around the DX magnet and the ZDC.

The threshold dependence of the selection of charged particle candidates was 
also studied. 
The change in charged particle fraction was less than 1\% so that 
the threshold dependence was negligibly small. 
Therefore, the factor 1.5 was a reasonable estimate for the fraction of 
charged candidates between the data and simulation. 

\begin{figure}[htb]
\includegraphics[width=1.0\linewidth]{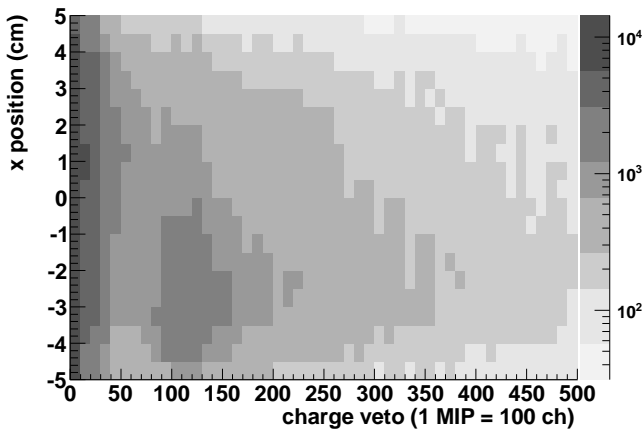}
\caption{
The $x$ position calculated by the SMD $vs.$ the charge distribution in the 
charge veto counter for the experimental data.
Most charged events were distributed in the negative $x$ region which is the direction 
of beam bending by the DX magnet. 
}
\label{fig_m_21:2}
\end{figure}

The proton background was estimated and included in the systematic 
uncertainties. 
According to the simulation study for the structure of proton events, 
proton events should be detected in the direction of beam bending which 
is negative $x$ for the south ZDC. 
This behavior was confirmed by the experimental data as shown in 
Fig.~\ref{fig_m_21:2} which is a plot of the $x$ position determined by 
the SMD $vs.$ the charge distribution in the charge veto counter. 
Most charged candidates were distributed in the negative $x$ region. 
We assumed the proton background $A_N$ equals zero and evaluated its 
systematic uncertainty by the dilution method with $A_{N}^{bg}$=0 in 
the asymmetry analysis. 

\section{Cross section measurement}

\subsection{Analysis}\label{sec_ax_1}

The differential cross section with respect to $x_F$ was measured:
\begin{eqnarray}
  \frac{d\sigma}{dx_F} = \frac{N_{{\rm neutron}}}{\mathcal{L}} \frac{1}{dx_F}, \label{eq_ax_1}
\end{eqnarray}
where $N_{{\rm neutron}}$ is the number of neutrons after the correction of cut efficiencies and the energy unfolding.

For the cross section analysis, 6.5 million events taken by the ZDC trigger 
were used from the sampled luminosity of 240 nb$^{-1}$. 
The acceptance cut $r$$<$2 cm was used to select kinematics similar to 
the ISR experiment. 
We assumed the beam axis was the same as the ZDC center in this analysis and 
the deviation was evaluated as a systematic uncertainty. 
The beam axis compared to the ZDC center is discussed in Appendix A. 

\begin{figure}[htb]
\includegraphics[width=1.0\linewidth]{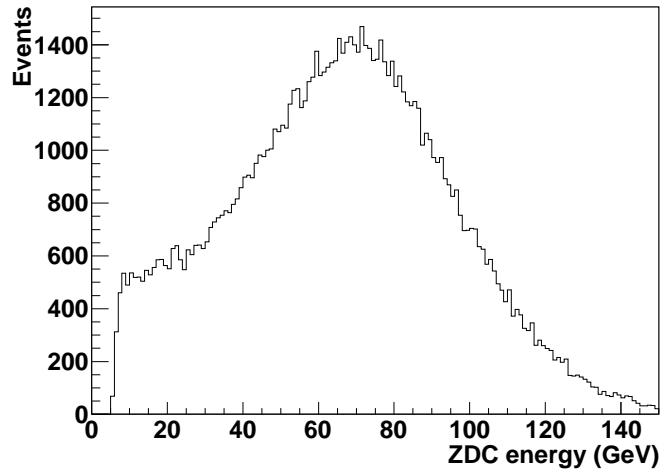}
\caption{The energy distribution measured with the ZDC after the neutron identification and the acceptance cut ($r$$<$2 cm, corresponds to $p_T$$<$0.11 GeV/$c$). }
\label{fig_ax_1}
\end{figure}
 
Figure \ref{fig_ax_1} shows the energy distribution measured with the ZDC after the 
neutron identification and the acceptance cut.
The energy spectrum was peaked at about 70 GeV, and this was used 
for a stability check of the ZDC gain run by run, which was found to be stable.
The ratio of the neutron yield to the BBC counts without the collision vertex requirement
was used for a stability check of the neutron selection, and 
it was also found to be stable.

One background source was beam-gas interaction. 
Beam-gas events are normally removed by requiring a forward-backward
coincidence of the BBC detectors.  However, this could not be done for 
the ZDC triggered events. 
Instead, we evaluated the fraction of beam-gas background using the 9 
noncolliding bunch crossings with the combination of 
filled and empty bunches at PHENIX. 
We found that the fraction was 0.0062$\pm$0.0004 on average, negligibly 
small. 

The neutron hit position was calculated by the centroid method using 
the distribution of scintillator charge above the threshold in the SMD, 
Eq.~(\ref{eq_e_1}). 
In this analysis, the same threshold was applied to the data and simulation 
and the efficiency of the SMD cut was estimated by simulation. 
The difference of efficiencies caused by uncertainty of the SMD cut 
efficiency was estimated using the nearly pure neutron sample by the neutron 
identification with the second ZDC cut (section \ref{sec_m_9}). 
The energy spectrum was corrected based on the SMD cut efficiency before 
the energy unfolding. 

 The measured neutron energy with the ZDC is smeared 
by the energy resolution.
 To extract the initial energy distribution, 
it is necessary to unfold the measured energy distribution.
 The energy unfolding method is described in Appendix B.

The ZDC energy response to neutrons below 100 GeV was 
found by the simulation to be nonlinear as described in section \ref{sec_m_3}.
 This nonlinearity was included in the transition matrix $A$ of Appendix B,
and corrected by the energy unfolding.
 Since the hadronic interaction could only be determined from simulation, a systematic 
uncertainty was included, using the variation of the cross section evaluated 
with a different matrix $A$ with a linear response.

 The efficiency of the experimental cuts, 
including the neutron identification and the acceptance cut, 
for the unfolded $x_F$ distribution was estimated 
by the simulation with the single neutron event generator.
 The acceptance cut used the radius, $r$, 
and the efficiency was evaluated from the $p_T$ distribution based on 
Eq.~(\ref{eq_i_3}):
$p_T \approx x_F \cdot E_p \cdot \theta_n \approx 0.056 \cdot x_F \cdot 
r$ GeV/$c$.

\begin{figure}[htb]
\includegraphics[width=1.0\linewidth]{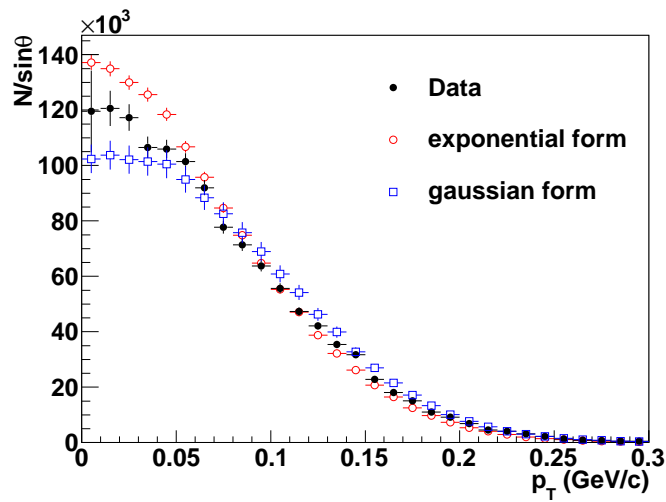}
\caption{(color online)
Comparison of the $p_T$ distribution from experimental data (black closed 
circles) and two simulations using Gaussian form (blue open squares) and 
exponential form (red open circles) inputs. 
}
\label{fig_ptdist}
\end{figure}

For the $p_T$ distribution, we used two distributions: a Gaussian 
form $d\sigma/dp_T \propto 
\exp(-a p_T^2)$, where $a(x_F)$ was obtained by HERA \cite{Chekanov:2007tv}
with error evaluation, and an exponential form 
$d\sigma/dp_T \propto \exp(-b p_T)$, where $b=4.8\pm 0.3$ (GeV$/c$)$^{-1}$ 
which was used in the ISR analysis \cite{Engler:1974nz,Flauger:1976ju}.
The simulated $p_T$ distributions with those two input
distribution were compared with experimental data normalized to 
the same total entries.
It was found that the
differences between data and those two inputs were not
large as shown in Fig.~\ref{fig_ptdist}.

\begin{figure}[htb]
\includegraphics[width=1.0\linewidth]{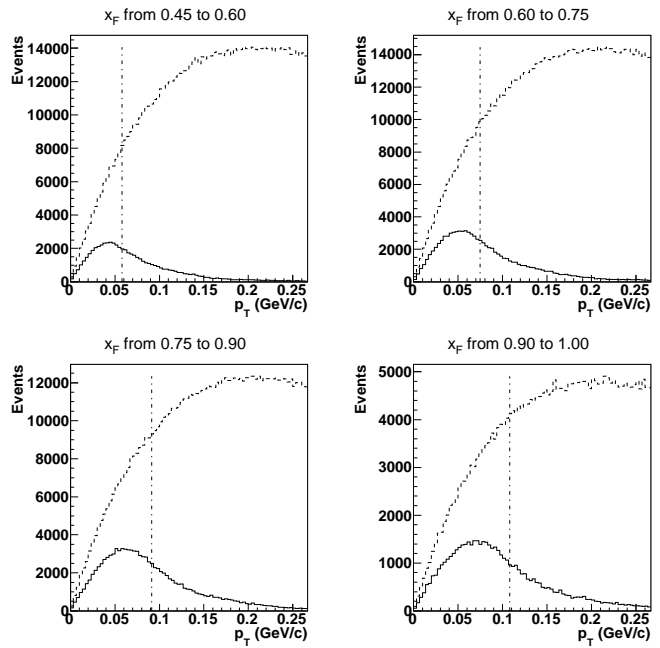}
\caption{ 
Simulated $p_T$ distributions using the Gaussian $p_T$ shape. 
Initial $p_T$ distributions are shown as dashed lines in each $x_F$ region. 
The expected $p_T$ region for the acceptance, $r<2$ cm, is below 
the vertical dot-dashed lines, which are the maximum $p_T$ calculated as 
$ \approx 0.11 \cdot x_F$ GeV/$c$. 
The actual $p_T$ distributions with the experimental cuts are shown as 
solid lines. 
}
\label{fig_ax_10}
\end{figure}

Figure \ref{fig_ax_10} shows the simulated $p_T$ distributions (dashed line) 
in each $x_F$ bin.
The geometrical maximum $p_T$ for the acceptance cut, $r<2$ cm, in each $x_F$ 
are given by 
$p_T^{\rm Max}=0.11 \cdot x_F$ GeV/$c$, shown as dot-dashed vertical lines.
The actual $p_T$ distributions with the experimental cuts
were smeared due to the position resolution and the 
energy resolution, shown as solid lines.
Ratios of these counts are the efficiency for the 
experimental cuts, and are listed in Table \ref{tab_ax_2}.
The errors were derived considering the uncertainty in
the parameter $a(x_F)$ in the Gaussian form evaluated by HERA.
There is no significant difference in the result in case of using the ISR
(exponential) $p_T$ distribution. 

\begin{table}[htb]
\caption{
The expected $p_T$ for $r<$2 cm, mean $p_T$ value with the 
experimental cut, 
and the efficiency for the experimental cut estimated by the 
simulation (Fig.~\ref{fig_ax_10}).
The errors were derived considering the uncertainty in
the parameter $a(x_F)$ in the Gaussian form evaluated by HERA.
}
\begin{ruledtabular}\begin{tabular}{lcc}
neutron    & mean      & efficiency \\
$x_F$      & $p_T$     &            \\
           & (GeV/$c$) &            \\
\hline
0.45--0.60 & 0.072     & 0.779 $\pm$ 0.014 (1.8\%) \\
0.60--0.75 & 0.085     & 0.750 $\pm$ 0.009 (1.2\%) \\
0.75--0.90 & 0.096     & 0.723 $\pm$ 0.006 (0.8\%) \\
0.90--1.00 & 0.104     & 0.680 $\pm$ 0.016 (2.3\%)
\end{tabular}\end{ruledtabular}
\label{tab_ax_2}
\end{table}

 The mean values of the simulated $p_T$ distributions in 
each energy region are also listed in Table \ref{tab_ax_2}.
 The cross section was obtained after the correction 
of the energy unfolding and the cut efficiency.

\begin{table}[htb]
\caption{
Systematic uncertainties for the cross section measurement.
The absolute normalization error is not included in these errors.
The absolute normalization uncertainty was estimated by BBC counts to be 
9.7\% (22.9$\pm$2.2 mb for the BBC
trigger cross section).
}
\begin{ruledtabular}\begin{tabular}{lcc}
                   & exponential $p_T$ form & Gaussian $p_T$ form \\
\hline
$p_T$ distribution & 3 -- 10\%              & 7 -- 22\% \\
beam center shift  & \multicolumn{2}{c}{3 -- 31\%} \\
proton background  & \multicolumn{2}{c}{3.6\%} \\
multiple hit       & \multicolumn{2}{c}{7\%} \\
\\
total              & 11 -- 33\%             & 16 -- 39\%
\end{tabular}\end{ruledtabular}
\label{table:syserr_cs}
\end{table}

Table \ref{table:syserr_cs} summarizes all systematic uncertainties evaluated as
the ratio of the variation to the final cross section values.
The absolute normalization error is not included in these errors.
It was estimated by BBC counts to be 9.7\% (22.9$\pm$2.2 mb for the BBC
trigger cross section).

The background contamination in the measured neutron energy with the ZDC 
energy from 20 to 140 GeV for the acceptance cut of $r<$ 2 cm was estimated 
by the simulation with the {\sc pythia}  event generator. 
The background from protons was estimated to be 2.4\% in the simulation. 
The systematic uncertainty in the experimental data was determined to be 
1.5 times larger than this as discussed in section \ref{sec_m_9}. 
Multiple particle detection in each collision was estimated to be 7\% 
with the $r<$ 2cm cut. 

In the cross section analysis, we evaluated the beam center shift described 
in Appendix A as a systematic uncertainty. 
For the evaluation, cross sections were calculated in the different 
acceptances according to the result of the beam center shift while requiring 
$r$$<$2 cm, and the variations were applied as a systematic uncertainty. 

\subsection{Result}\label{sec_re_1}

\begin{table}[htb]
\caption{
The result of the differential cross section $d\sigma/dx_F\ ({\rm mb})$
for neutron production in $p$$+$$p$ collisions at $\sqrt{s}$=200 GeV.
The first uncertainty is statistical, after the unfolding, and the second
is the systematic uncertainty.
The absolute normalization error, 9.7\%, is not included.
}
\begin{ruledtabular}\begin{tabular}{lcc}
$\langle x_F \rangle$ & exponential $p_T$ form    & Gaussian $p_T$ form \\
\hline
0.53                  & 0.243$\pm$0.024$\pm$0.043 & 0.194$\pm$0.021$\pm$0.037 \\
0.68                  & 0.491$\pm$0.039$\pm$0.052 & 0.455$\pm$0.036$\pm$0.085 \\
0.83                  & 0.680$\pm$0.044$\pm$0.094 & 0.612$\pm$0.044$\pm$0.096 \\
0.93                  & 0.334$\pm$0.035$\pm$0.111 & 0.319$\pm$0.037$\pm$0.123
\end{tabular}\end{ruledtabular}
\label{tab_re_1}
\end{table}

\begin{figure}[htb]
\includegraphics[width=1.0\linewidth]{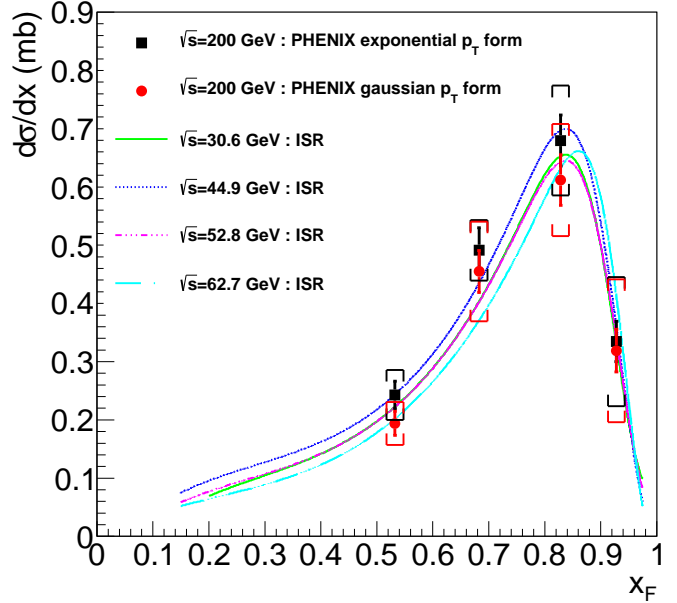}
\caption{(color online)
The cross section results for forward neutron production in $p$$+$$p$ collisions 
at $\sqrt{s}$=200 GeV are shown. 
Two different forms, exponential (squares) and Gaussian (circles), were used 
for the $p_T$ distribution. 
Statistical uncertainties are shown as error bars for each point, and 
systematic uncertainties are shown as brackets. 
The integrated $p_T$ region for each bin is $0<p_T<0.11x_F$ GeV/$c$. 
Shapes of ISR results are also shown. 
Absolute normalization errors for the PHENIX and ISR are 9.7\% and 20\%, 
respectively. 
}
\label{fig_re_1:2}
\end{figure}

The differential cross section, $d\sigma/dx_{F}$, for forward neutron 
production in $p$$+$$p$ collisions at $\sqrt{s}$=200 GeV was determined using 
two $p_T$ distributions: a Gaussian form, as used in HERA analysis, and 
an exponential form, used for ISR data analysis. 
The results are listed in Table \ref{tab_re_1} and plotted in 
Fig.~\ref{fig_re_1:2}. 
We show the results for $x_F$ above 0.45 since the data below 0.45 are 
significantly affected by the energy cut-off before the unfolding. 
The $p_T$ range in each $x_F$ bin is $0<p_T<0.11x_F$ GeV/$c$ 
from Eq.~(\ref{eq_i_2}) with the acceptance cut of $r<2$ cm. 
The absolute normalization uncertainty for the PHENIX measurement, 9.7\%, 
is not included. 

 Invariant cross sections measured at the ISR 
experiment were converted to differential cross sections 
for the comparison with the PHENIX data.
 The conversion formula from the invariant cross section
$E d^3\sigma/dp^3$ to $d\sigma/dx_F$ is described 
with the approximation in the forward kinematics as
\begin{eqnarray}
\frac{d\sigma}{dx_F} = \frac{2\pi}{x_F} \int_{Acc.} 
E\frac{d^3 \sigma}{d^3 p} p_T dp_T, \label{eq15}
\end{eqnarray}
 where $Acc.$ means the $p_T$ range of the PHENIX 
acceptance cut; $0<p_T<0.11x_F$ GeV/$c$ for the $r<2$ cm cut.
 As a $p_T$ shape, we used an exponential form $\exp(-4.8p_T)$ 
which was obtained from the $0.3<x_F<0.7$ region from the ISR results 
\cite{Engler:1974nz,Flauger:1976ju}.

For both the table and figure, we give the PHENIX results for two
$p_T$ shapes, the exponential shape used for the ISR results, and
the Gaussian shape used for HERA results.

The measured cross section at $\sqrt{s}$=200 GeV is consistent with the ISR result,
 indicating that $x_{F}$ scaling is satisfied at the higher center of mass energy.
 This result is consistent with the OPE model.

\section{Single transverse spin asymmetry measurement}

\subsection{Analysis}\label{sec_aa_1}

The single transverse spin asymmetry is obtained from the azimuthal 
modulation of neutron production relative to the polarization direction of 
a transversely polarized beam on an unpolarized target, and normalized by 
an independent measurement of the beam polarization.
The stable polarization direction of protons is vertical with respect to 
the accelerator plane. 
There is an approximately equal number of bunches filled with the spin of 
polarization-up protons as of polarization-down protons.
With both beams polarized, single-spin analyses were performed by 
taking into account the polarization states of one beam, averaging over 
those of the other.
The beam polarizations were measured using fast carbon target 
polarimeters \cite{Jinnouchi} at a different location at RHIC with
several measurements in each fill.  The carbon target measurements were
normalized to absolute polarization 
measurements made by a separate polarized atomic hydrogen jet polarimeter 
\cite{Okada:2005gu}. 
The polarizations ranged from 0.43 to 0.48 for the Blue beam and from 0.46 
to 0.52 for the Yellow beam. 
Systematic uncertainty for the Blue beam polarization is 5.9\%, and that 
for the Yellow beam polarization is 6.2\%. 

\begin{figure}[htb]
\includegraphics[width=0.91\linewidth]{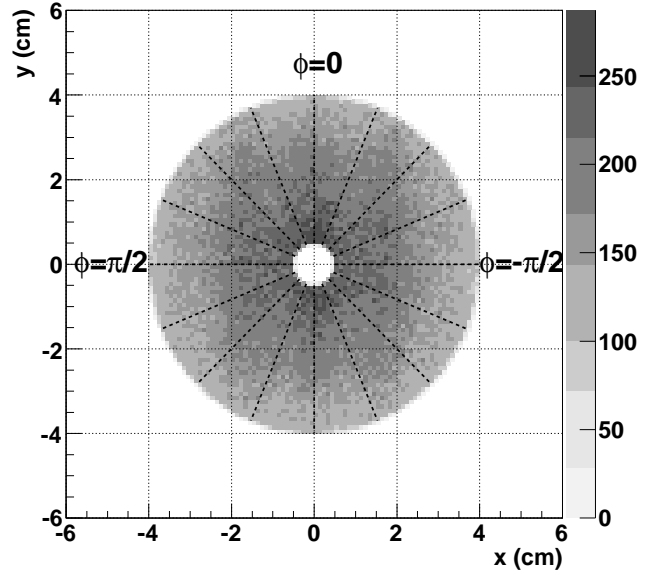}
\caption{ 
  The acceptance definition for the $\phi$ dependence of $\epsilon_N$,
shown as a plot of the measured neutron position at the ZDC.
  The acceptance was divided into 16 slices in a radial pattern and the 
asymmetry was calculated by the square root formula starting at $\phi = -\pi/2$ to $\pi/2$.
}
\label{fig_aa_4}
\end{figure}

The acceptance definition for the azimuthal angle ($\phi$) of
neutron production is shown in Fig.~\ref{fig_aa_4}, where 
the polarization-up direction points to $\phi=0$.
The acceptance cut at the ZDC required 0.5$<$$r$$<$4.0 cm.
The acceptance area was divided into 16 slices in a radial pattern.
For the asymmetry calculation, we used a square-root formula which cancels 
many systematic uncertainties,
 such as detector and luminosity asymmetries:
 \begin{equation}
   \epsilon_{N}(\phi)
   = \frac{\sqrt{N^{\ua}_{\phi} N^{\da}_{\phi+\pi}}
     - \sqrt{N^{\ua}_{\phi+\pi} N^{\da}_{\phi}}}
      {\sqrt{N^{\ua}_{\phi} N^{\da}_{\phi+\pi}}
     + \sqrt{N^{\ua}_{\phi+\pi} N^{\da}_{\phi}}}.
       \label{eq_ip_3:1}
 \end{equation}
where $N^{\ua}_{\phi}$ ($N^{\da}_{\phi}$) is the number of events with 
polarization-up (-down) producing neutrons to azimuthal angle $\phi$.

A correction $C_{\phi}$ is applied, discussed later, to account for smearing from
position resolution.
After normalization by the polarization, $P$, we obtain the asymmetry as,
\begin{eqnarray}
  {\cal A}(\phi) = \frac{1}{P} \frac{1}{C_{\phi}} \epsilon_{N}({\phi}). \label{eq_ip_3:2}
\end{eqnarray}

For this analysis, we used 6.5 million and 17.6 million events for the ZDC 
trigger sample and ZDC$\otimes$BBC trigger sample respectively from 
the sampled luminosity of 240 nb$^{-1}$. 
A ZDC energy cut was required to select 40--120 GeV in the measured energy.

The raw measured asymmetry $\epsilon_N(\phi)$ divided by the polarization 
are fitted to a sine:
\begin{eqnarray}
  {\cal A}(\phi) = A_N\ {\rm sin}(\phi-\phi_0), \label{eq_ip_3:3}
\end{eqnarray}
where $\phi_0$ allows a deviation of the maximum asymmetry axis from vertical.

In the present analysis, we used only the south ZDC detector, which 
faces the Yellow beam. 
The forward neutron asymmetry uses the polarized Yellow beam and sums 
over the polarization states of the Blue beam bunches.  Following the Basel 
(Ann Arbor) Convention \cite{Ashkin:1977ek}, a positive $A_N$ indicates 
more production to the left of the polarized (Yellow) beam, for 
the polarization-up bunches in the Yellow beam.
The asymmetry for neutrons produced backward was measured using a polarized 
Blue beam, summing over the polarization states of the Yellow beam bunches. 
In order to follow the Basel Convention, signs of the backward $A_N$ were 
inverted from the fitting results. 
A positive $A_N$ would indicate more neutron production to the left of 
the Blue (polarized) beam for polarization-up bunches. 

We performed two sets of simulations to estimate the smearing parameters, 
$C_{\phi}$, which were correlated to the neutron energy-dependent position 
resolution (section \ref{sec_m_3}).
The energy distributions for the simulation inputs were determined in 
the same way as the cross section analysis (Section \ref{sec_ax_1}). 

The $\epsilon_N(\phi)$ was smeared from the ${\cal A}(\phi)$ due to 
position resolution.  From Eq.~(\ref{eq_ip_3:2}), the smearing 
parameter, $C_{\phi}$, can be evaluated from simulation as,

\begin{eqnarray}
  C_{\phi} = \frac{ A_N^{Output} }{ A_N^{Input} }, \label{eq31.1}  
\end{eqnarray}

\noindent where $A_{N}^{Output}$ corresponds to the 
$\epsilon_N(\phi)$ of the experimental data; it includes effects of 
the experimental cut and the position resolution. As 
$A_{N}^{Input}$, we generated neutrons with the sine modulated 
${\cal A}(\phi)$ as Eq.~(\ref{eq_ip_3:3}) with 
$A_N$=$A_N^{Input}$=$-$0.10. The smeared amplitude was obtained as 
$A_N^{Output}$=$-$0.076 and their ratio, 0.76, is the correction 
factor of the smearing effect, $C_{\phi}$ = 0.760 $\pm$ 0.015 (ZDC 
trigger). For the ZDC$\otimes$BBC trigger we obtained the smearing 
parameter $C_{\phi}$ = 0.746 $\pm$ 0.016 (ZDC$\otimes$BBC trigger).

For the analysis of the $x_F$ dependence of $A_N$, we chose bins of 
40-60, 60-80, and 80-120 GeV
in the measured ZDC energy.
Events with ZDC energy greater than 120 GeV were eliminated from this analysis
(3.8\% of the events).
 Similar simulations and calculations of $C_{\phi}$ were performed for the analysis of
the $x_F$ dependence of the asymmetry with both the ZDC trigger and
ZDC$\otimes$BBC trigger.

 After correction for the smearing effect, we obtain the measured energy dependence of $A_N$.
 The mean $x_F$ values for the ZDC trigger sample and ZDC$\otimes$BBC trigger sample were evaluated by the simulations
 which were modified to reproduce the measured energy distributions for each trigger sample.

The background contamination was studied by the simulation with 
the {\sc pythia}  event generator. 
In the analysis of the $x_F$ dependence of $A_N$, an acceptance cut of 
$r<$ 3 cm was applied. 

 After the neutron identification and the acceptance cut, as described in 
section \ref{sec_m_9}, the neutron purities were 0.975 $\pm$ 0.006 for the 
ZDC trigger sample, and 0.977 $\pm$ 0.010 for ZDC$\otimes$BBC trigger sample.
 Main background contributions were the $K^{0}$ and proton.
 According to the discussion in section \ref{sec_m_9}, we applied the systematic 
uncertainty contributed from the proton only.
 They were 1.4\% and 1.0\% for the ZDC trigger and ZDC$\otimes$BBC trigger 
respectively and were increased by the factor 1.5 estimated higher frequency
of proton background in the experimental data, compared to simulation, to give
 2.1\% and 1.5\%, which were included as systematic uncertainties.
 Multiple particle detection in each collision was estimated to be 6.5\%
for the ZDC trigger and 5.9\% for the ZDC$\otimes$BBC trigger for the 0.5$<$$r$$<$4.0 cm cut.

To evaluate the systematic uncertainty for determination of the beam axis, 
$A_N$ were calculated with center positions as ($x$,$y$) = (0.46, 0.00), 
(0.00, -1.10) and  (0.46, -1.10) cm while keeping the acceptance cut, 
0.5$<$$r$$<$4.0 cm. 
These values were chosen based on measurements of the beam center as 
discussed in Appendix A. 
Maximum variations to final values, which were calculated by ($x$,$y$) 
= (0.00, 0.00) cm, were included as systematic uncertainties. 

 Since the smearing effect was caused by the position resolution, the systematic uncertainty 
of the position resolution, 14\% (section \ref{sec_m_3p}), should be reflected in the 
uncertainties for the result.
 This was evaluated with a variation of the asymmetry calculated with 14\% 
increased position resolution uncertainty in the simulation.
 The asymmetry was reduced by 4.2\%.
 This was assigned as a systematic uncertainty of the smearing correction. 

A technique called ``bunch shuffling'' was used to check for systematic 
effects in the asymmetry measurements due to a variation of beam 
characteristics bunch by bunch. 
By randomly assigning bunch polarization directions, we create data sets 
of experimental data with little or no net polarization, and compare 
the resulting measured asymmetry with statistical uncertainties. 
The fluctuation of measured asymmetries should correspond to
the statistical uncertainty. 
We concluded that the fake asymmetry from bunch characteristics is less 
than 0.39$\sigma_{stat}$ and 0.36$\sigma_{stat}$ for the ZDC trigger 
and ZDC$\otimes$BBC trigger respectively. 
We do not include these uncertainties in the final systematic uncertainties 
for $A_N$. 

$p_T$-correlated uncertainties from the beam center shift were evaluated 
in a similar way to the cross section analysis described in Appendix A. 
They were 0.004 in the $x_F$-integrated analysis, and 0.004--0.010 in the 
$x_F$-dependent analysis.

\begin{table}[htb]
\caption{Scale uncertainties for the $A_N$ measurements.}
\begin{ruledtabular}\begin{tabular}{lcc}
                   & ZDC trigger & ZDC$\otimes$BBC trigger \\
\hline
proton background  & 2.1\%       & 1.5\% \\
multiple hit       & 6.5\%       & 5.9\% \\
smearing           & \multicolumn{2}{c}{4.2\%} \\
\\
total              & 8.0\%       & 7.4\%
\end{tabular}\end{ruledtabular}
\label{table:syserr_an}
\end{table}

Scale uncertainties are summarized in Table \ref{table:syserr_an} for 
the $A_N$ measurements. 
Values are presented as scale variations to the final values. 
Total uncertainties were calculated by quadratic sum. 
The scale uncertainty from the beam polarization is not included in 
the table. 
The uncertainty in the Yellow beam polarization which was used in 
the forward neutron asymmetry measurement was $\pm$ 6.2\%, and that in 
the Blue beam polarization used in the backward neutron asymmetry 
measurement was $\pm$ 5.9\%. 

\subsection{Azimuthal modulation of forward neutron production}\label{sec_re_2}

In this section we present the results for the azimuthal modulations for 
neutron production, within the acceptance from $\theta_n=0.3$ mrad 
($r=0.5$ cm at ZDC) to $\theta_n=2.2$ mrad ($r=4$ cm), and ZDC energy 
from 40 GeV to 120 GeV. 

\begin{figure}[htb]
\includegraphics[width=1.0\linewidth]{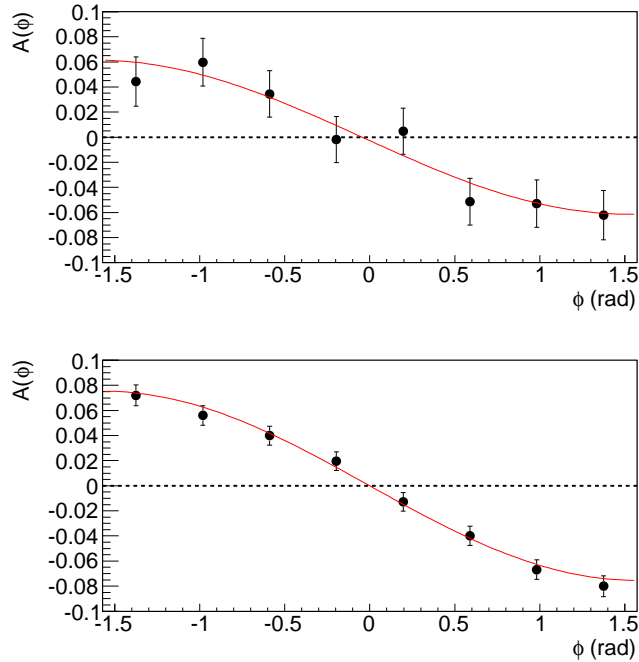}
\caption{(color online)
Results for the azimuthal modulation for forward neutron production from 
polarized $p$$+$$p$ collisions at $\sqrt{s}$=200 GeV in the ZDC trigger sample 
(top) and the ZDC$\otimes$BBC trigger sample (bottom).
The error bars indicate the statistical uncertainties. 
Results for a ${\rm sin}(\phi)$ fit to the data are indicated. 
The $p_T$-correlated systematic uncertainties from the beam center shift, 
and scale uncertainties listed in Table \ref{table:syserr_an} and 
polarization scale uncertainties are not included. 
}
\label{fig_re_3_1}
\end{figure}

\begin{figure}[htb]
\includegraphics[width=1.0\linewidth]{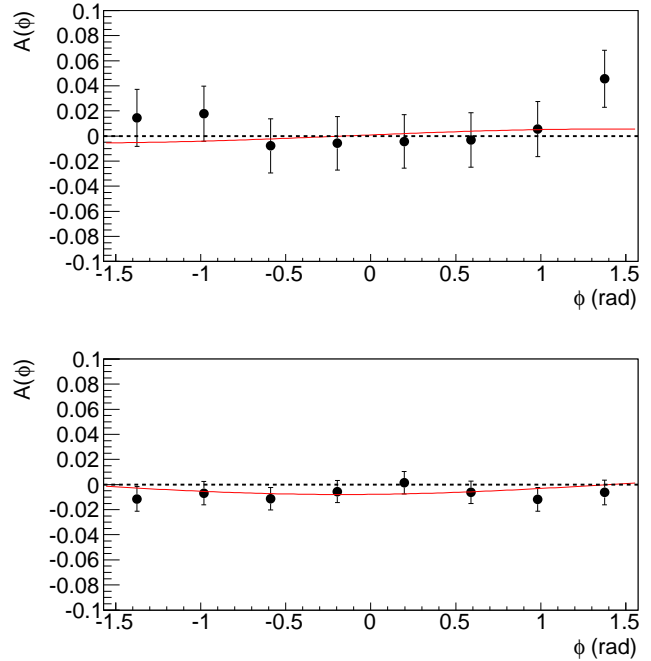}
\caption{(color online)
Results for the azimuthal modulation for backward neutron production from 
polarized $p$$+$$p$ collisions at $\sqrt{s}$=200 GeV in the ZDC trigger sample 
(top) and the ZDC$\otimes$BBC trigger sample (bottom). 
The error bars indicate the statistical uncertainties. 
Results for a ${\rm sin}(\phi)$ fit to the data are indicated. 
The $p_T$-correlated systematic uncertainties from the beam center shift, 
and scale uncertainties listed in Table \ref{table:syserr_an} and 
polarization scale uncertainties are not included. 
}
\label{fig_re_3_2}
\end{figure}

Asymmetries ${\cal A}(\phi)$ were calculated for eight azimuthal angle bins, 
using Eq.~(\ref{eq_ip_3:1}) and Eq.~(\ref{eq_ip_3:2}). 
Figure \ref{fig_re_3_1} and \ref{fig_re_3_2} present ${\cal A}(\phi)$ for 
the two trigger conditions, for forward and backward neutron production 
respectively. 
Statistical uncertainties are shown in the figure. 
The $p_T$-correlated systematic uncertainties from the beam center shift are 
not shown. 
In addition, there are scale uncertainties listed in Table 
\ref{table:syserr_an} and polarization scale uncertainties. 

A significant asymmetry is present for forward neutron production. 
The ${\cal A}(\phi)$ data were fitted with a sine curve, 
Eq.(\ref{eq_ip_3:3}), to obtain $A_N$. 
The azimuthal offsets, $\phi_0$, were consistent with $\phi_0=0$. 
The results obtained for $A_N$ are: 
$A_N = -0.061 \pm 0.010 (stat) \pm 0.004 (syst)$ ($\chi^2/ndf = 3.05/6$) 
for the ZDC trigger sample and 
$A_N = -0.075 \pm 0.004 (stat) \pm 0.004 (syst)$ ($\chi^2/ndf = 2.22/6$) 
for the ZDC$\otimes$BBC trigger sample. 
There is no observed asymmetry for backward neutron production. 
The results for backward neutron production for $A_N$ are: 
$A_N = -0.006 \pm 0.011 (stat) \pm 0.004 (syst)$ ($\chi^2/ndf = 5.18/6$) 
for the ZDC trigger sample and 
$A_N = -0.008 \pm 0.005 (stat) \pm 0.004 (syst)$ ($\chi^2/ndf = 3.31/6$) 
for the ZDC$\otimes$BBC trigger sample. 

To compare with the previous result \cite{Fukao:2006vd} from the 
polarimeter development experiment at RHIC, we compared to the $A_N$ of 
the forward 
ZDC$\otimes$BBC trigger sample. 
The amplitude of the measured $A_N$ was; 
$A_N = (-0.090\pm0.006\pm0.009)\times(1.00^{+0.52}_{-0.25})$. 
Errors indicate the statistics, systematics and the scaling uncertainty 
from the polarization measurement. 
The two results are consistent within the errors, including the scaling 
uncertainties. 
We note that the two measurements used slightly different detection 
coverages for the charged particle interaction trigger: 
2.2$<$$|\eta|$$<$3.9 in the horizontal and vertical directions for 
the polarimeter development experiment, and  3.0$<$$|\eta|$$<$3.9 for the 
PHENIX experiment. 

\subsection{$x_F$ dependence of $A_N$}\label{sec_re_3}

\begin{table}[htb]
\caption{
The results of the $x_F$ dependence of $A_N$ for neutron
production in the ZDC trigger sample of $p$$+$$p$ collisions at $\sqrt{s}$=200 GeV.
First and second uncertainties show statistical and $p_T$-correlated 
systematic uncertainties, respectively.
Scale uncertainties from the asymmetry measurements and the beam polarization 
are not included. 
}
\begin{ruledtabular}\begin{tabular}{lcc}
$\langle x_F \rangle$ & $A_N$                         & $\chi^2/ndf$ \\
\hline
-0.776                & -0.0059$\pm$0.0252$\pm$0.0095 & 11.6/7 \\
-0.682                & -0.0219$\pm$0.0255$\pm$0.0035 & 6.833/7 \\
-0.568                & -0.0050$\pm$0.0303$\pm$0.0076 & 9.252/7 \\
 0.568                & -0.0503$\pm$0.0263$\pm$0.0076 & 7.012/7 \\
 0.682                & -0.0625$\pm$0.0221$\pm$0.0035 & 2.68/7 \\
 0.776                & -0.0772$\pm$0.0217$\pm$0.0095 & 5.38/7 
\end{tabular}\end{ruledtabular}
\label{tab_re_4}
\end{table}

\begin{table}[htb]
\caption{
The results of the $x_F$ dependence of $A_N$ for neutron
production in the ZDC$\otimes$BBC trigger sample of $p$$+$$p$ collision at
$\sqrt{s}$=200 GeV.
First and second uncertainties show statistical and $p_T$-correlated 
systematic uncertainties, respectively.
Scale uncertainties from the asymmetry measurements and the beam polarization 
are not included. 
}
\begin{ruledtabular}\begin{tabular}{lcc}
$\langle x_F \rangle$ & $A_N$                         & $\chi^2/ndf$ \\
\hline
-0.749                &  0.0035$\pm$0.0117$\pm$0.0082 & 2.672/7 \\
-0.664                & -0.0093$\pm$0.0106$\pm$0.0037 & 2.915/7 \\
-0.547                & -0.0033$\pm$0.0115$\pm$0.0096 & 6.783/7 \\
 0.547                & -0.0629$\pm$0.0097$\pm$0.0096 & 13.27/7 \\
 0.664                & -0.0657$\pm$0.0090$\pm$0.0037 & 5.425/7 \\
 0.749                & -0.0667$\pm$0.0099$\pm$0.0082 & 5.003/7 
\end{tabular}\end{ruledtabular}
\label{tab_re_5}
\end{table}

\begin{figure}[htb]
\includegraphics[width=1.0\linewidth]{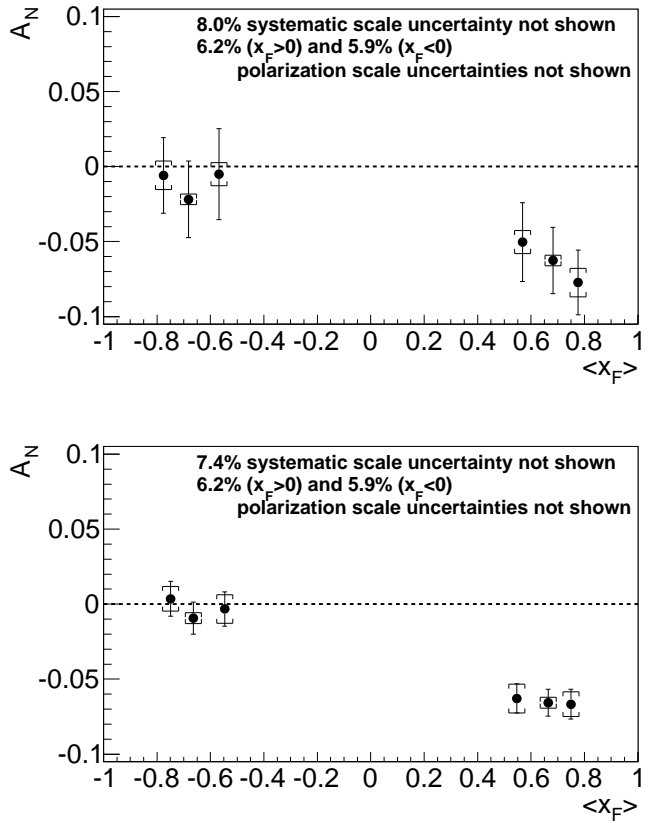}
\caption{
The $x_F$ dependence of $A_N$ for neutron production in the ZDC trigger 
sample (top) and for the ZDC$\otimes$BBC trigger sample (bottom). 
The error bars show statistical uncertainties and brackets show 
$p_T$-correlated systematic uncertainties. 
Systematic scale uncertainties listed in Table \ref{table:syserr_an} and 
polarization scale uncertainties are not included. 
}
\label{fig_re_4}
\end{figure}

The $x_{F}$ dependence of $A_N$ for production is listed in 
Table \ref{tab_re_4} and \ref{tab_re_5}, and plotted in Fig.~\ref{fig_re_4}. 
The ${\cal A}(\phi)$ data were fitted with a sine curve, Eq.(\ref{eq_ip_3:3}), 
to obtain $A_N$ with $\phi_0=0$. 
The mean $x_F$ values were determined according to section \ref{sec_aa_1}. 
Statistical uncertainties are shown as error bars and $p_T$-correlated 
systematic uncertainties are shown as brackets. 
Scale uncertainties from the asymmetry measurements and the beam polarization 
are not included. 
  
 We observe significant negative $A_N$ for neutron production 
in the positive $x_F$ region and 
with no energy dependence within the uncertainties, both for inclusive
neutron production and for production including a beam-beam interaction
requirement.  No significant backward neutron production asymmetry is observed.

\section{Discussion}\label{sec_dis_0}

The measurement of the cross section for the $p$$+$$p$ production of neutrons 
at $\sqrt{s}$=200 GeV has been presented here, and it is consistent with 
$x_F$ scaling when compared to ISR results. 
These cross sections are described by the OPE model in Regge calculus 
\cite{Capella:1975qk,Kopeliovich:1996iw,Nikolaev:1997cn,Nikolaev:1998se,
D'Alesio:1998bf,Kaidalov:2006cw,Bunyatyan:2006vy}. 
Therefore, the observed large asymmetry for neutron production at RHIC, as 
presented in \cite{Fukao:2006vd} and here, may arise from the interference 
between a spin-flip amplitude due to the pion exchange and nonflip 
amplitudes from other Reggeon exchanges. 
So far our knowledge of Reggeon exchange components for neutron production 
is limited to the pion. 
Under the OPE interpretation, the asymmetry has sensitivity to 
the contribution of all spin nonflip Reggeon exchanges, even if 
the amplitudes are small. 
Recently Kopeliovich {\it et al.} calculated the $A_N$ of forward 
neutron production from the interference of pion and Reggeon exchanges, 
and the results were in good agreement with the PHENIX data 
\cite{Kopeliovich:2011bx}. 

We can also discuss our results based on the meson-cloud model 
\cite{meson-cloud}. 
This model gives a good description for the result from a Drell-Yan 
experiment at FNAL, E866\cite{Towell:2001nh}. 
In this model, the Drell-Yan process is generated by the interaction 
between the $d$ quark in one proton and the $\bar{d}$ quark in 
the $\pi^{+}$ of $p \rightarrow n \pi^{+}$ state for other proton. 
In this model the neutron should be generated with very forward kinematics, 
possibly similar to the kinematics of the results presented here. 
The meson-cloud model was successfully applied to neutron production 
in the ISR experiment \cite{D'Alesio:1998bf} and we expect it is applicable
to our $A_N$ and cross section measurements for higher energy $p$$+$$p$ 
collisions.

\section{Conclusion} \label{sec_conc_0}

We have measured the cross section and single transverse spin 
asymmetry, $A_N$, for very forward neutron production in polarized 
$p$+$p$ collisions at $\sqrt{s}$=200 GeV.  The results from the PHENIX 
experiment at RHIC were based on a zero degree hadronic calorimeter 
(ZDC) augmented by a shower maximum detector, covering neutron 
production angles to $\theta_n$=2.2 mrad.  A large $A_N$ for neutron 
production had been observed in a polarimeter development experiment at 
RHIC, using an electromagnetic calorimeter to identify neutrons, with 
coarse neutron energy resolution\cite{Fukao:2006vd}. The PHENIX 
experiment then outfitted existing ZDC detectors to act as polarimeters 
to monitor the beam polarizations and polarization directions at the 
experiment.  The results presented here are based on studies with the 
ZDC polarimeter, which due to a much better measurement of the neutron 
energy, provide first measurements of the neutron production cross 
section at RHIC energy, and the dependence of $A_N$ on the neutron 
energy.

The measured cross section is consistent with $x_F$ scaling from ISR 
results. Within uncertainties, the observed $A_N$ were consistent with 
the previous result at RHIC~\cite{Fukao:2006vd} and for $x_F > 0.45$ 
(the region measured by this experiment) no significant $x_F$ 
dependence was observed.  We also present measured $A_N$ for neutrons 
produced backward from the polarized beam.  These results are 
consistent with zero.

The cross sections for large $x_F$ neutron production, as well as those 
in $e+p$ collisions at HERA, are largely reproduced by a one pion exchange model (OPE).
Using this model, the observed large asymmetry for the neutron production would be 
considered to come from the interference between a spin-flip amplitude 
due to the pion exchange and nonflip amplitudes from other Reggeon exchanges.
On the basis of the OPE model, the large neutron $A_N$ would have 
sensitivity to the contribution of other Reggeon exchanges.

Future measurements of neutron production cross sections and asymmetries 
will include analysis of RHIC runs at $\sqrt{s}$=62.4 GeV and at 500 GeV. 
The measurements at different center of mass energies will probe the $x_F$ 
and $p_T$ dependence for neutron production at fixed, very forward
production angles $\theta_n<$2.2 mrad. 
 
\section*{ACKNOWLEDGMENTS}

We thank the staff of the Collider-Accelerator and Physics
Departments at Brookhaven National Laboratory and the staff of
the other PHENIX participating institutions for their vital
contributions.  We acknowledge support from the 
Office of Nuclear Physics in the
Office of Science of the Department of Energy,
the National Science Foundation, 
a sponsored research grant from Renaissance Technologies LLC, 
Abilene Christian University Research Council, 
Research Foundation of SUNY, 
and Dean of the College of Arts and Sciences, Vanderbilt University 
(U.S.A),
Ministry of Education, Culture, Sports, Science, and Technology
and the Japan Society for the Promotion of Science (Japan),
Conselho Nacional de Desenvolvimento Cient\'{\i}fico e
Tecnol{\'o}gico and Funda\c c{\~a}o de Amparo {\`a} Pesquisa do
Estado de S{\~a}o Paulo (Brazil),
Natural Science Foundation of China (P.~R.~China),
Ministry of Education, Youth and Sports (Czech Republic),
Centre National de la Recherche Scientifique, Commissariat
{\`a} l'{\'E}nergie Atomique, and Institut National de Physique
Nucl{\'e}aire et de Physique des Particules (France),
Bundesministerium f\"ur Bildung und Forschung, Deutscher
Akademischer Austausch Dienst, and Alexander von Humboldt Stiftung (Germany),
Hungarian National Science Fund, OTKA (Hungary), 
Department of Atomic Energy and Department of Science and Technology (India),
Israel Science Foundation (Israel), 
National Research Foundation and WCU program of the 
Ministry Education Science and Technology (Korea),
Ministry of Education and Science, Russian Academy of Sciences,
Federal Agency of Atomic Energy (Russia),
VR and Wallenberg Foundation (Sweden), 
the U.S. Civilian Research and Development Foundation for the
Independent States of the Former Soviet Union, 
the US-Hungarian Fulbright Foundation for Educational Exchange,
and the US-Israel Binational Science Foundation.

\appendix

\section{The study of the beam axis on the detector geometry}

 The ZDC center was aligned to the beam axis at the beginning of the 2003 run.
 We assumed that the beam axis was on the ZDC center in this analysis of 2005 data.  We used
two approaches to estimate the beam and ZDC alignment.  Peripheral neutrons
from a heavy ion run just prior to the polarized proton run gave centers of
$x = 0.28\pm0.01$ cm and $y = -0.07\pm0.01$ cm at the south ZDC.  The center of
the asymmetry $A_N$ was also used to determine the center of the beam axis at
the ZDC, since $A_N$ must be zero at zero production angle.  
We used the ZDC$\otimes$BBC trigger sample in this analysis. 
The asymmetry was measured for a vertically polarized beam to obtain the center in $x$, and for a
special run with horizontally polarized beam to obtain the center in $y$.  The
results were $x = +0.46\pm0.08$ cm and $y = -1.10\pm0.14$ cm.  The results of the
two techniques agreed reasonably for $x$, and did not agree for $y$.

The beam axis shifts that we observed were considered as systematic 
uncertainties for the results.
The uncertainties were determined from variations of the cross section and asymmetry 
obtained by moving the center of acceptance while keeping the same cut 
region (for example, $r$$<$2 cm for the cross section analysis).

\section{Energy Unfolding} 

The measured neutron energy with the ZDC is smeared by the energy 
resolution. 
For the extraction of the original energy distribution, it is necessary 
to unfold the measured energy distribution. 
We use an unfolding method proposed in \cite{Blobel:2002pu}. 

We assume that the initial distribution $x(E)$ is smeared to the 
measured distribution $y(E')$ and this smearing is described by 
a linear combination. 
Their relation can be given by a transition matrix $A(E',E)$ as, 
\begin{eqnarray} 
y(E') = A(E',E)x(E). \label{eq11} 
\end{eqnarray} 
or $\vec{y} = A \vec{x}$. 

\begin{figure}[thb] 
 \includegraphics[width=1.0\linewidth]{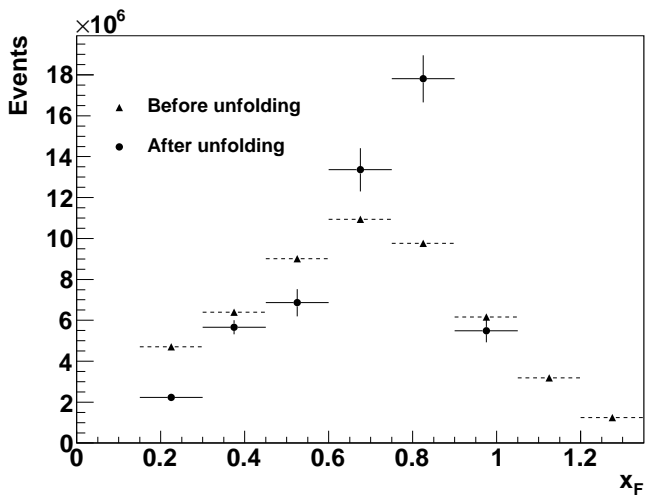} 
\caption{Energy distributions before and after the energy unfolding. 
 The unfolding was performed so that statistics was conserved. 
} 
\label{fig_ax_15} 
\end{figure} 
   
If the smearing effect is large, the result is very sensitive to 
a small change of $A$. 
It can be discussed using an orthogonal decomposition. 
The matrix $A$ is diagonalized into $D$ with a transformation matrix 
$U$,
\begin{eqnarray} 
\vec{c} = D \vec{b}, \label{eq24} 
\end{eqnarray} 
where $D = U^{-1} A U$, and $\vec{c}=U^{-1}\vec{y}$ and 
$\vec{b}=U^{-1}\vec{x}$ are new vectors transformed from $\vec{y}$ 
and $\vec{x}$, respectively. 
The diagonal elements of the matrix $D$ are the eigenvalues 
$\lambda_{j}$ of the matrix $A$. 
Each of the coefficients $b_j$ and $c_j$ in $\vec{c}=D\vec{b}$ is 
transformed independently of any other coefficient by using 
eigenvalue $\lambda_j$, 
\begin{eqnarray} 
c_j = \lambda_j \cdot b_j. \label{eq25} 
\end{eqnarray} 

In order to perform the unfolding, the coefficients $c_j$ have been 
affected by statistical fluctuations of the elements of measured 
vector $\vec{y}$. 
The $b_j$ which includes the information of initial vector $x$ is 
obtained by $b_j = c_j / \lambda_j$. 
The statistical fluctuation of the $c_j$ amplified
in the case of small eigenvalue $\lambda_j$, resulting in instability.
Reasonable result can be obtained by cutting the $c_j$ which has 
a large statistical uncertainty. 

First, the coefficients $c_j$ were calculated. 
Three sets of the transition matrix $A$, which have the same energy 
resolution but different initial energy distributions, were prepared 
with a simulation to check the statistical error propagation of the 
$c_j$. 
Initial shapes were prepared to increase, be flat and decrease as 
a function of $x_F$. 
These shapes are close to the cross sections at $p_T$ $\approx$ 
0.0 GeV/$c$, 0.2 GeV/$c$ and 0.4 GeV/$c$ in the ISR results. 
Energy spectra before and after the unfolding are plotted in 
Fig.~\ref{fig_ax_15}.
Horizontal axis is changed to $x_F$ by Eq.~(\ref{eq_i_2}).

 

\end{document}